\documentclass[conference]{IEEEtran}

\usepackage{graphicx}
\usepackage{amsmath, amssymb}
\usepackage{algorithm2e}
\RestyleAlgo{ruled} % To get lines above and below the algorithm
\SetAlgoCaptionSeparator{.}
\SetKwInput{KwInput}{Input}
\SetKwInput{KwOutput}{Output}

\usepackage{caption}
\usepackage{multirow}
\usepackage{multicol}
\usepackage{xcolor}
\usepackage{enumitem}
\usepackage{mathtools}
\usepackage{comment}

\def\BibTeX{{\rm B\kern-.05em{\sc i\kern-.025em b}\kern-.08em
    T\kern-.1667em\lower.7ex\hbox{E}\kern-.125emX}}

\begin{document}

% TITLE
\title{DL-AMC: Deep Learning for Automatic Modulation Classification}

% AUTHORS
\author{
\IEEEauthorblockN{Faheem Ur Rehman\IEEEauthorrefmark{1}, Qamar Abbas\IEEEauthorrefmark{2}, and M. Karam Shehzad\IEEEauthorrefmark{3}}
\IEEEauthorblockA{\IEEEauthorrefmark{1}Faculty of Management Sciences, FAST National University of Computer and Emerging Sciences, Islamabad, Pakistan\\
Email: faheem.urrehman@isb.nu.edu.pk}
\IEEEauthorblockA{\IEEEauthorrefmark{2}Faculty of Computer Science and Engineering, GIK Institute of Engineering Sciences and Technology, Pakistan\\
Email: qamar.abbas@giki.edu.pk}
\IEEEauthorblockA{\IEEEauthorrefmark{3}Radio Interface and Access Group, Nokia Standards, Paris, France\\
Email: muhammad-karam.shehzad@nokia.com}
}

\maketitle
\begin{abstract}
Automatic Modulation Classification (AMC) is a
signal processing technique widely used at the physical layer
of wireless systems to enhance spectrum utilization efficiency.
In this work, we propose a fast and accurate AMC system, termed DL-AMC, which leverages deep learning techniques. Specifically, DL-AMC is built using convolutional neural network (CNN) architectures, including ResNet-18, ResNet-50, and MobileNetv2. To evaluate its performance, we curated a comprehensive dataset containing various modulation schemes. Each modulation type was transformed into an eye diagram, with signal-to-noise ratio (SNR) values ranging from -20 dB to 30 dB. We trained the CNN models on this dataset to enable them to learn the discriminative features of each modulation class effectively.
Experimental results show that the proposed DL-AMC models achieve high classification accuracy, especially in low SNR conditions. These results highlight the robustness and efficacy of DL-AMC in accurately classifying modulations in challenging wireless environments.
\end{abstract}
\section{Introduction}
\IEEEPARstart{A}utomatic Modulation Classification (AMC) is a technique used to identify the modulation type of an incoming signal while accounting for factors such as fading, noise, and channel-induced interference. AMC applies to both cooperative and non-cooperative communication, involving tasks such as channel tracking, as well as estimating and detecting signal parameters \cite{leblebici2024cnn}. With the rapid advancements in Deep Learning (DL) techniques, researchers have increasingly explored the application of a convolutional neural network (CNN) for AMC. 

Over the past decade, numerous machine learning (ML) techniques have
been developed to classify unknown signals and identify their modulation type. To achieve effective modulation classification under real-world channel conditions, \cite{huang2024generalized} proposes a novel feature-based method, called EVM-AMC (Error Vector Magnitude-based Automatic Modulation Classification), which outperforms traditional approaches in terms of robustness and generalization. Furthermore, \cite{fu2021lightweight} introduces a decentralized learning approach (DecentAMC) which leverages model consolidation and a lightweight separable convolutional neural network (S-CNN) architecture to significantly enhance training efficiency and reduce communication overhead. These strategies have become increasingly popular in the field of signal processing and have largely replaced traditional methods, such as Support Vector Machines (SVMs) \cite{mohsen2024automatic} and Artificial Neural Networks (ANN) \cite{popoola2013effect}, \cite{popoola2011novel}.
In \cite{popoola2013effect}, the authors have selected eight modulation types that exhibit high performance even at low SNR values. In \cite{popoola2011novel}, a multilayer perceptron (MLP) was used to accurately classify twelve different modulation types across various SNR levels. The adoption of DL approaches has further enhanced feature learning from complex, high-dimensional data. The
CNN architecture utilized in \cite{liu2017deep} achieves an accuracy of 83.8\% at high SNR, Meanwhile, \cite{o2016convolutional} presents a CNN specifically designed for classifying ten different modulation schemes. 

One of the limitations of the earlier research \cite{li2019signal,
hu2018robust,
zhang2018automatic,
west2017deep} is their reliance on a small set of modulation schemes. Furthermore, these studies disregard the importance of signal preprocessing.
In \cite{li2019signal}, the authors employ a Deep Belief Network (DBN) trained on four modulation schemes, achieving 10\% accuracy at the lowest SNR. Meanwhile, in \cite{hu2018robust}, the authors utilize a recurrent neural network (RNN) \cite{shehzad2022ml} with four modulation schemes, achieving a 48\% accuracy under the lowest SNR conditions. \cite{zhang2018automatic} introduces the use of long short-term memory (LSTM) \cite{shehzad2022real} with eleven modulation schemes, attaining a 12\% accuracy at the lowest SNR. Similarly, \cite{west2017deep} adopts Convolutional Long Short-Term Deep Neural Networks (CLDNN) with eleven modulation schemes, yielding a 12\% accuracy at the lowest SNR. In contrast, we propose an innovative approach that surpasses traditional methods by preprocessing incoming signals and generating dynamic visual representations. This enables us to clearly illustrate the impact of noise on system performance, offering valuable insights that are directly applicable to real-world systems. Our methodology not only improves understanding but also facilitates more informed decision-making in environments where noise interference plays a critical role. The main contributions of our work are outlined below.
\begin{itemize}
\item DL-AMC introduces a novel transformation of complex signals into eye diagrams, which provide a more robust representation than traditional constellation diagrams. 
Unlike constellation diagrams, eye diagrams inherently capture temporal variations, jitter, and noise effects, enhancing the resilience of DL models in low-SNR conditions.
\item We develop an innovative signal representation that enables DL frameworks to efficiently classify modulation schemes. Our approach includes three comprehensive DL models, surpassing traditional algorithms in accuracy and robustness.
\item DL-AMC carefully fine-tunes the input layer dimensions and optimizes the output layer configuration to align with the unique characteristics of our dataset. These strategic adjustments enhance feature extraction, leading to improved accuracy and computational efficiency. The proposed approach demonstrates strong potential for real-world modulation classification applications.
\end{itemize}
The rest of the paper is organized as follows: Section\,\ref{meth} explains the proposed DL-AMC architecture.  Section\,\ref{sec3} presents the results, and Section\,\ref{con} concludes the paper.
%%%%%%%%%%%%
\renewcommand{\figurename}{Fig.}
\begin{figure*}[ht]
   \centering
\includegraphics[width=1.0\textwidth]{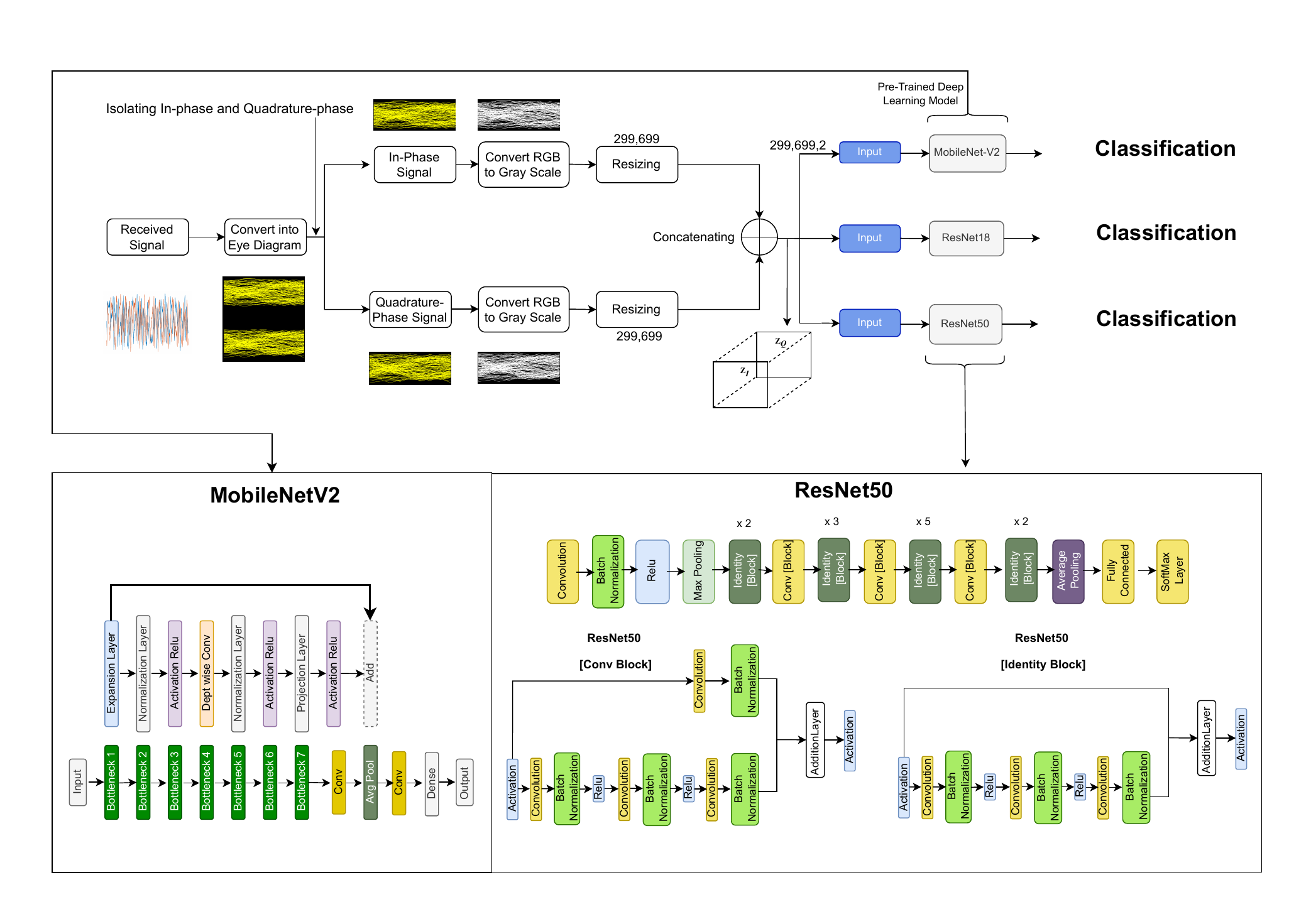}
   {\captionsetup{font={small,rm}, labelsep=period, singlelinecheck=false, skip=5pt}
   \caption{DL-AMC workflow diagram. This diagram illustrates the proposed system's workflow, starting with the reception of a modulated signal. It undergoes processing to create an eye diagram, which is split into $I$ and $Q$ components, converted to grayscale, resized, and then concatenated for unified representation, and fed to DL models, offering an overview of the system's signal processing stages.}
   \label{workflow}}
\end{figure*}
\section{DL-AMC}\label{meth}
DL-AMC aims to precisely classify modulation schemes by preprocessing the input signals, allowing the DL model to efficiently learn their features. In the following subsection, we discuss DL-AMC in detail.
%%%%%%%%%%%%
\subsection{System Overview}
\label{sys}
The proposed model aims to classify the modulation techniques with high accuracy. The flow diagram shown in Fig.\,\ref{workflow}
illustrates the overall methodology\footnote{Three DL models ResNet18, ResNet50, and MobileNetv2-V2 are employed. Since ResNet50 is an advanced variant of ResNet18, we only displayed the results of ResNet18 due to space limitations.
}. A breakdown of the different components and their connections is as follows:
\begin{itemize}
    \item \textbf{Receive Signal Block:} This block represents the input signal received by the system, which is assumed to be a modulated signal transmitted over a wireless channel.
    \item \textbf{Eye Diagram Block:}
    The received signal is processed to generate an eye diagram of every frame plotting $N_s$ samples in each trace of the modulation class $i = 1, 2, \ldots, M\: \text{and corresponding frames }\: j = 1, 2, \ldots, N$, where $M$ is the number modulation classes and $N$ corresponds to several frames in $i^{th}$ modulation class. An eye diagram of $i^{th}$ modulation class is created as
   \begin{equation}
   \mathbf{A}_{i} = \sum_{j=1}^{N} \text{eyediagram}[(i,j), N_s].
 \end{equation}
    \item \textbf{Inphase and Quadrature Phase Split:}
    The eye diagram is split into two paths: one for the in-phase,
    \textit{I}
    component and the other for the quadrature-phase, \textit{Q} component. This separation is typically done for signal processing purposes, as modulation schemes often involve the simultaneous transmission of two orthogonal components.

To enhance data preparation for deep learning (DL) models, applying a cropping function to the eye diagram is advisable. A cropping function in this context is used to remove unnecessary elements, such as the axes, that may be included when saving the eye diagram. This step ensures that only the relevant portion of the diagram is preserved, focusing on the signal itself without any additional, non-informative elements. Removing the axes prevents the introduction of unnecessary linearity or redundant features into the dataset, which could otherwise impact the performance of the DL model. By cropping out these extraneous elements, we ensure that the data is cleaner and better suited for model training and analysis.
    \item \textbf{RGB to Gray Scale Block: } Both \textit{I} and \textit{Q} phase signals are processed individually in the RGB to Gray Scale block. This conversion is performed to transform the signals from the RGB color space to a single grayscale channel. This simplifies further processing and analysis.
    \item \textbf{Resize Block:} The grayscale signals are then resized to a standardized or desired dimension given as

     \begin{equation}
 \mathbf{B}_{i,\textit{I}} = \sum_{k=1}^K \text{resize}[\text{RGB2Gray}{\{\text{crop}}(\mathbf{A}_{ik}^\textit{I})\}]\:,
       \end{equation}

\begin{equation}
 \mathbf{B}_{i,\textit{Q}} = \sum_{k=1}^K \text{resize}[\text{RGB2Gray}{\{\text{crop}}(\mathbf{A}_{ik}^\textit{Q})\}]\:,
       \end{equation}
    where $K=\text{length}(\mathbf{A})$.
    \item \textbf{Concatenation:}
    The resized \textit{I} and \textit{Q} phase signals are concatenated to form a unified representation of the received signal, represented as
  \begin{equation}
 \mathbf{Z}_{\textit{I}} = \{ \mathbf{B}_{ik,\textit{I}}\}\:,
\end{equation}
  \begin{equation}
   \mathbf{Z}_{\textit{Q}} = \{ \mathbf{B}_{ik,\textit{Q}} \}\:.
  \end{equation}
\end{itemize} 

In a nutshell, the block diagram outlines the signal processing steps involved in the system, starting from the received signal and going through the stages of eye diagram generation, signal separation, color space conversion, resizing, and finally, the concatenation of the processed signals. This processing pipeline prepares the signal for subsequent stages, such as modulation classification using CNNs. The
preprocessing steps are summarized in Algorithm \ref{algorithm1}.
\subsection{Deep Learning Models}

Besides efficient preprocessing, our research focuses on developing and enhancing DL models while optimizing hyperparameter tuning to improve modulation classification. We employed three DNN models: ResNet50, ResNet18\footnote{We also trained other DL models, but their accuracy was suboptimal. Our focus is not on addressing the limitations of previous models.}, and MobileNetv2. To make these models adaptable to our dataset, we modified the input layer size from (224, 224, 3) to (299, 699, 2), and adjusted the output softmax layer to correspond to the number of modulation types, denoted as $M$. This adjustment was necessary because our dataset represents signals as eye diagrams, which are intrinsically wider and require a different input format.
To accommodate variations in input size and adapt the output layer, we redesigned the aforementioned models, ensuring optimal feature extraction and classification performance across diverse signal conditions. The main reason for changing the input size from (224, 224, 3) to (299, 699, 2) is the unique properties of eye diagrams. These diagrams contain more detailed temporal information and are organized around two channels, one for \textit{I} phase data and the other for \textit{Q} phase data. However, it demonstrates the effective application of hyperparameter tuning to tailor existing architectures for a specialized task, thereby making a significant contribution to the field of modulation classification.\\
In the context of DL, the optimizer used for training the DL model is stochastic gradient descent with momentum (SGDM). This optimization algorithm is known for efficiently navigating complex parameter spaces by combining the benefits of gradient descent with a momentum term. The learning rate for the optimization process is set at 0.001, determining the step size during parameter updates. The training process spans a maximum of 20 epochs, representing the number of times the entire dataset is passed through the model. Each training iteration involves processing batches of data, with a minimum batch size set at 32. Our DL model is trained across various SNR levels, which are transformed into eye diagrams to effectively capture signal distortion. This approach enables the model to classify signals under different conditions while reducing sensitivity to inference patterns, allowing it to learn complex signal characteristics more effectively. Additionally, the model can be fine-tuned with newly observed inference patterns, ensuring adaptability in dynamic environments.
\begin{algorithm}[ht]
\SetAlgoLined
\DontPrintSemicolon
\For{$i = 1$ \textbf{to} M}
{
\For{$j = 1$ \textbf{to} N} 
    {

{$\mathbf{A}_{ij}$$\gets$eyediagram[($i,j$),$N_s$]}}

\For{$k = 1$ \textbf{to} $\mathbf{K}$}  
{{$\mathbf{B}_{ik,\textit{I}}$$\gets$crop($\mathbf{A}_{ik}$) }\newline
{$\mathbf{B}_{ik,\textit{I}}$$\gets$RGB2Gray({$\mathbf{B}_{ik,\textit{I}}$}) }
\newline
{$\mathbf{B}_{ik,\textit{I}}$$\gets$resize({$\mathbf{B}_{ik,\textit{I}}$)} }

{$\mathbf{B}_{ik,\textit{Q}}$$\gets$crop($\mathbf{A}_{ik}$) }

{$\mathbf{B}_{ik,\textit{Q}}$$\gets$RGB2Gray({$\mathbf{B}_{ik,\textit{Q}}$)} }

{$\mathbf{B}_{ik,\textit{Q}}$$\gets$resize({$\mathbf{B}_{ik,\textit{Q}}$)} }
}        

{$\mathbf{Z}_{I}$$\gets$$\mathbf{B}_{\textit{ik,I}}$} 

{$\mathbf{Z}_{Q}$$\gets$$\mathbf{B}_{\textit{ik,Q}}$}}
\caption{Preprocessing Procedure}
\label{algorithm1}
 \end{algorithm}
\subsection{Dataset Generation}
The dataset encompasses a diverse range of modulation schemes\footnote{B-FM, AM-SSB, AM-DSB, 16-QAM, 64-QAM, 128-QAM, 256-QAM, QPSK, BPSK, 8-PSK, 16PSK, GFSK, GPFSK, 4-PAM.}. 
Across all SNR levels, a consistent number of $5\times10^3$ frames were generated per modulation scheme, resulting in a total of $70\times10^3$ frames for each SNR value.
The dataset's final size is defined by a cumulative frame count of $420\times10^3$  frames, highlighting the extensive range of signal variations captured. This substantial dataset facilitates comprehensive evaluation and optimization of modulation classification models, ensuring robustness across diverse signal conditions. This dataset serves as a valuable resource for evaluating and refining modulation classification models across diverse SNR conditions. By encompassing a broad range of signal variations, it supports the development of robust DL frameworks, ultimately advancing signal processing techniques and enhancing the analysis of modern communication systems. The dataset is divided as: 70\% for training, 20\% for validation, and 10\% for test.
The test dataset comprises modulation frames across all SNR levels, ensuring comprehensive performance evaluation. Each modulation scheme is sampled at a rate of 200 kHz with 1024 samples per frame. For digital modulation, a center frequency of 902 MHz is assumed, while analog modulation frames are centered at 100 MHz. The synthetic signals used for training are impaired with additive white Gaussian noise (AWGN) at a different SNR, Rician multipath fading modeled with path delays of [0, 1.8, 3.4] samples and corresponding path gains of [0, -2, -10] dB. These impairments are applied independently to each frame to ensure the model generalizes well across different interference conditions. The frame-wise processing approach enables the CNN to classify signals without explicit dependence on the number of users, making it scalable for multi-user environments.
\section{Result and Evaluation}\label{sec3}
\begin{figure}[h!]
    \centering
\includegraphics[width=0.47\textwidth]{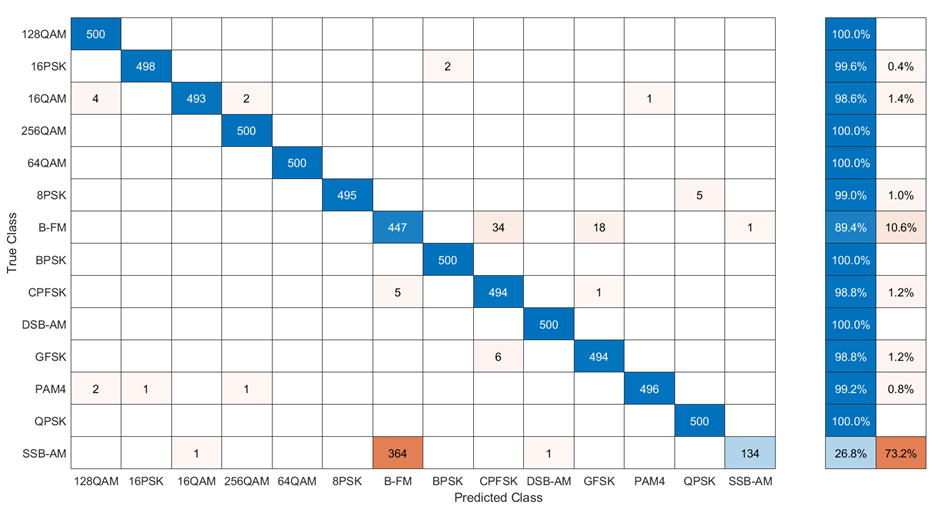}
    \captionsetup{font={small,rm}, labelsep=period, singlelinecheck=false, skip=5pt}
    \caption{MobileNetv2 model with $-20\,$dB.}
    \label{m1}
\end{figure}

\begin{figure}[h!]
    \centering
\includegraphics[width=0.47\textwidth]{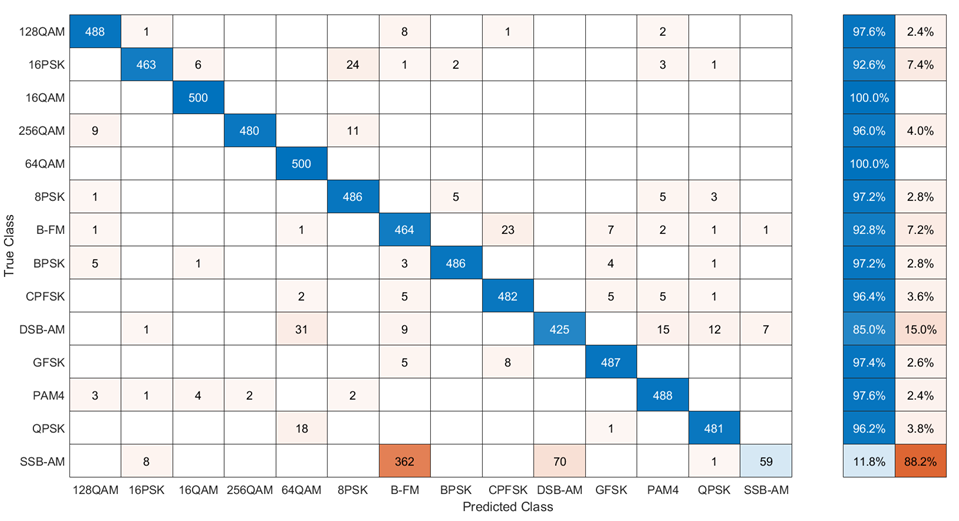}
\captionsetup{font={small,rm}, labelsep=period, singlelinecheck=false, skip=5pt}
    \caption{ResNet18 model with $-20\,$dB.}
    \label{m2}
\end{figure}

\begin{figure}[h!]
    \centering
\includegraphics[width=0.47\textwidth]{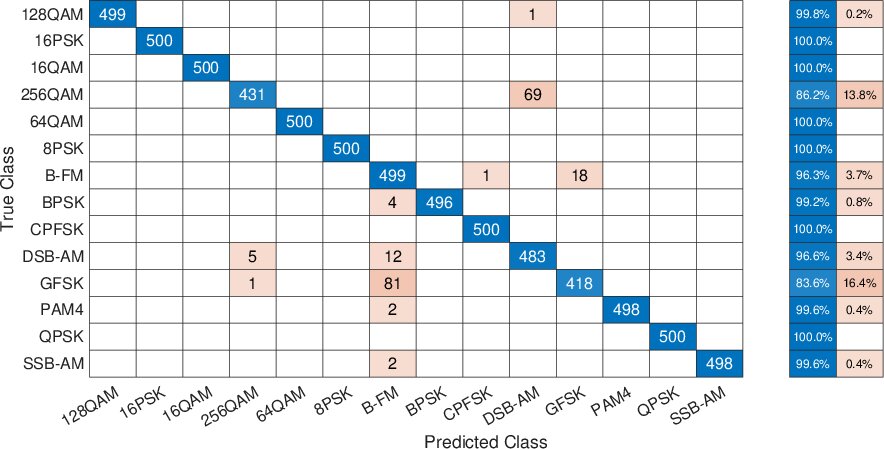}
\captionsetup{font={small,rm}, labelsep=period,singlelinecheck=false, skip=5pt}
    \caption{ResNet50 model with $-20\,$dB.}
    \label{m3}
\end{figure}

The DL models are evaluated using test data with varying SNR, $M=14$ and $N_s=8$. Through experiments, we found the lower SNR at which the models attained maximum accuracy. The confusion matrix for varying SNR ratio is presented in Figs. \ref{m1}, \ref{m2}, \ref{m3}.  
Due to space constraints, we present the confusion matrix specifically for the lowest SNR value ($-20\,$dB) to highlight the model's performance under extreme noise conditions.

It is important to note that even though only lower SNR values were shown, the three models ResNet-50, ResNet-18, and MobileNetv2 were tested across the complete range of SNR (dB) levels, covering $\{-20\,, -10\,, 0\,, 10\,, 20\,, 30\}$. We selected ResNet-18 and ResNet-50 due to their well-established effectiveness in feature extraction for complex visual patterns. The deeper architecture of ResNet-50 enables learning richer hierarchical representations, which enhances performance, particularly in low-SNR conditions. 

In summary, we have observed that ResNet-50 showed the best classification accuracy across all SNRs. ResNet-18 and MobileNetv2 gave good accuracy, with occasional changes in SNR level. All three DL models lose classification accuracy as SNR drops, suggesting signal quality loss. On the other side, ResNet-50 is the most noise-resistant neural network, followed by ResNet-18 and MobileNetv2. 

\subsection{Comparison with State-of-the-Art}
We rigorously tested our DL models against leading state-of-the-art solutions -- DBN \cite{li2019signal}, RNN \cite{hu2018robust}, LSTM \cite{zhang2018automatic}, and CLDNN \cite{west2017deep} -- across a range of SNR. The results demonstrated that our approach consistently outperforms these models. For example, at $-20$ dB, our model achieves an accuracy of $93.6\%$, while DBN, RNN, LSTM, and CLDNN show accuracies of only $10\%, 48\%, 12\%,$ and $12\%,$ respectively, even at higher SNR values. This stark contrast highlights the robustness of our model, particularly at lower SNRs. The key factor driving this exceptional performance is the innovative data processing methodology we propose, which enhances the model's ability to handle noise effectively.
\label{res}
\renewcommand{\figurename}{Fig.}
\section{Conclusion}
\label{con}
We have observed that DL-based Automatic Modulation Classification (DL-AMC) significantly enhances performance compared to state-of-the-art solutions, demonstrating improved robustness and accuracy in diverse signal conditions. DL-AMC involved a transformative step, converting complex signals into informative eye diagrams, which have proven to be a more resilient representation than traditional constellation diagrams. Furthermore, a significant architectural modification to the DL model is implemented by fine-tuning the input layer dimensions and optimizing the output layer configuration. This strategic adjustment is driven by the unique characteristics of the dataset, particularly when visualizing signals as eye diagrams. The wider format enables the incorporation of precise signal features, enhancing the model's ability to capture intricate patterns. Collectively, these contributions refine the preprocessing approach and tailor the DL model to better interpret dataset complexities, ultimately improving overall performance.

In the future, we plan to explore the proposed methodology using real-world data. Also, we aim to leverage Kolmogorov-Arnold networks (KAN) \cite{liu2024kan} to address catastrophic forgetting, enhancing the model's adaptability in dynamic environments. To further evaluate its performance, we will conduct real-world testing using actual received signals, ensuring its robustness and reliability in realistic communication system.
\bibliographystyle{IEEEtran}
\bibliography{references}
\end{document}